**Title**

- Experimental Observation of Extremely Strong Defect-Phonon Scatterings in Semiconductor Single Crystals

**Authors**

Zifeng Huang[1], Jianbo Liang[2], Yuxiang Wang[1], Zixuan Sun[1], Naoteru Shigekawa[2], Ming Li[1], Runsheng Wang[1], and Zhe Cheng[1,3,*]

**Affiliations**

[1] School of Integrated Circuits and Beijing Advanced Innovation Center for Integrated Circuits, Peking University, Beijing 100871, China.

[2] Department of Physics and Electronics, Osaka Metropolitan University, 1-1 Gakuen-cho, Naka-ku, Sakai, 599-8531, Japan.

[3] Frontiers Science Center for Nano-optoelectronics, Peking University, Beijing 100871, China.

[*]Author to whom correspondence should be addressed: zhe.cheng@pku.edu.cn

**Abstract**

The role of doping in tailoring thermal transport in semiconductors is critical for efficient thermal management in electronic devices. While the effects of doping have been extensively studied to tune electrical properties, its impact on thermal transport has not yet been thoroughly explored, particularly with respect to experimental investigations into exceptionally strong non-Rayleigh defect-phonon scattering phenomena. Herein, by combining the high-quality growth and advanced characterizations of cubic silicon carbide (3C-SiC) single crystals with well-controlled boron doping, we experimentally observe anomalous strong defect-phonon scatterings—among the strongest reported in common semiconductors—that exceeds the predictions of the classic mass-difference model by tens of times in magnitude. The measured thermal conductivity of doped 3C-SiC match excellently with those predicted by first-principle calculations in which resonant scattering of low frequency phonon is considered. Our findings not only shed light on the fundamental understanding of defect-phonon interactions and will also impact applications such as thermal management of electronics.

**Teaser**

Record-high thermal conductivity reduction of single crystal 3C-SiC was observed by boron dopants-induced resonant phonon scattering.

**MAIN TEXT**

**Introduction**

Doping, which intentionally introduces specific impurity atoms to supply additional electrons or holes, has been widely studied and extensively used to modulate carrier concentrations in semiconductors, facilitating the desired electrical properties for logic, memory, power, and optoelectronic devices(*1, 2*). Through precise control of carrier concentrations, doping has become a fundamental process in modern microelectronics. However, as electronic devices continue to shrink in size and increase in power density, thermal dissipation has emerged as a critical bottleneck, particularly for three-dimensional integrated circuits and high-power/high-frequency wide-bandgap semiconductor devices(*3, 4*). The thermal conductivity of semiconductors is fundamentally critical for the thermal management of micro- and nanostructured electronic devices, directly impacting their stability, reliability, and lifespan. While the effects of doping on electrical properties have been extensively studied, a unified and thorough understanding of doping-induced point defects and the consequent thermal conductivity reduction in semiconductors remains elusive due to the difficulties in sample growth, advanced thermal and structural characterizations.

According to classic kinetic theory, the thermal conductivity $k$ of a semiconductor is given by $k = \sum_i \frac{1}{3} C_i v_i^2 \tau_i$, where $C_i$, $v_i$, $\tau_i$ represent the heat capacity, group velocity, and relaxation time between two phonon scattering events for a specific polarization mode $i$(*5*). The total relaxation time $\tau_i$ is determined by contributions from different scattering mechanisms—such as phonon-phonon, phonon-defect, and phonon-electron interactions—following Matthiessen's rule: $\tau^{-1} = \sum_n \tau_n^{-1}$(*6*). In conventional doped semiconductors, substitutional defects are generally considered to induce Rayleigh scattering. Within this framework, the phonon-defect scattering relaxation time is proportional to the square of the mass difference between the host atom and the substituted atom(*7*). This scattering model has been successfully applied to predict the thermal conductivity of semiconductors such as 4H-SiC ,6H-SiC and GaN(*8, 9*). However, previous related researches suggest Rayleigh scattering alone may not fully account for all phonon-defect interactions. A thorough investigation of these effects is essential for a comprehensive understanding of thermal conduction in doped semiconductors.

In the 1960s, H. M. Rosenberg *et al*. observed an anomalously large reduction in thermal conductivity at low temperatures in hydrated zinc sulfate crystals doped with $Fe^{2+}$ ions compared to those doped with $Mg^{2+}$ ions, despite Mg exhibiting a much larger mass difference from the host Zn atom(*10*). He attributed it to phonon-spin scattering near a specific frequency, where the energy of a phonon induces the transition between different spin energy levels. Similarly, R. O. Pohl observed the interaction between phonons and in nonparamagnetic KCl crystal that led to a "dip" in thermal conductivity at low temperature, which cannot be explained by conventional Rayleigh scattering theory(*11*). They refer to this type of non-Rayleigh phonon scattering, involving interaction between spin and phonons near a certain frequency, as resonant phonon scattering, which significantly decrease the thermal conductivity of the host material. Similar scattering phenomenon has also been observed in thermoelectric materials, where weakly bound filler atoms inside cage-like structures induce resonant scattering(*12*).

While early studies by Pohl, Rosenberg, and others primarily focused on dielectric materials, the effect of phonon resonant scattering in doped semiconductors is of great importance. Several theoretical and experimental investigations(*13–18*), particularly on 3C-SiC—a key wide-bandgap semiconductor which possesses the second-highest thermal conductivity among large crystals (only surpassed by diamond)—have been reported(*14, 15, 17*). A 3C-SiC thin film has been shown to have significantly reduced thermal conductivity compared to high-purity 3C-SiC bulk crystal. However, the effects of impurities and boundary are not separated, making it difficult to understand the impact of defect-phonon scatterings alone(*17*). The thermal conductivity of polycrystalline 3C-SiC ceramic materials which are doped with nitrogen or co-doped with nitrogen and boron have also been studied(*19*). Nevertheless, previous theoretical and experimental results are controversial. For instance, in the theoretical study(*15*), the grain size used for fitting the theoretically calculated thermal conductivity (~0.8 μm) to the measured data of the 3C-SiC ceramic was nearly an order of magnitude smaller than the measured actual grain size (1–10 μm)(*19*). The strong grain boundary-phonon scatterings may compensate or couple with the defect-phonon scatterings. Additionally, Ivanova *et al.*'s report on nitrogen-boron co-doped 3C-SiC ceramics(*19*) is limited for defect analysis since it only provides carrier concentrations — a parameter that generally differs from defect concentrations, even in single-element-doped semiconductors. In the specific case of nitrogen-boron co-doped 3C-SiC, the carrier concentration would only represent the difference between activated nitrogen and boron dopants. However, as only a small fraction of boron dopants become activated at room temperature, substantial deviations in dopant concentrations are caused by relying solely on carrier concentration(*20*). Thus, separate quantification of nitrogen and boron defect concentrations requires additional advanced dopant characterization. Moreover, other structural imperfections may also scatter phonons due to the low quality of the 3C-SiC ceramic samples.

The limited availability of high-quality 3C-SiC single crystals with well-controlled doping levels, along with insufficient advanced structural and thermal characterizations, hinders the fundamental understanding of phonon scattering mechanisms—particularly the theory of resonant phonon scattering in 3C-SiC. In addition to well-designed sample growth, a detailed characterization of dopant distribution and its effect on phonon scattering is also crucial for a unified understanding of thermal conduction in doped semiconductors(*21*), as dopant clusters and non-uniform distribution may show up.

Here, we present advanced thermal and structural characterizations on high-quality boron-doped 3C-SiC single crystals to investigate defect-phonon scattering effects on thermal conductivity. By utilizing low-temperature chemical vapor deposition (LT-CVD) and time-domain thermoreflectance (TDTR), we synthesize high-quality 3C-SiC single-crystal samples and measure their thermal conductivity as functions of temperature and doping concentration. We also conduct comprehensive structural analysis using high-resolution scanning transmission electron microscopy (HR-STEM), Raman spectroscopy, and electron backscatter diffraction (EBSD) to characterize the crystal quality. Furthermore, secondary ion mass spectrometry (SIMS) and atom probe tomography (APT) are employed to quantify the concentrations and elemental distribution of dopants. These techniques collectively facilitate the direct comparison of experimental data with theoretical predictions to study the resonant phonon scattering mechanism.

## Results

### Structural characterization

Four 3C-SiC single crystals with intentionally controlled boron (B) doping are grown by LT-CVD. More details about the growth process can be found in the Materials and Methods section and ref(*17*). The B concentrations within the four crystals are determined by SIMS: ~$1\times10^{16}$ cm$^{-3}$ (non-doped, the detect limit of the SIMS technique), $6\times10^{18}$ cm$^{-3}$, $1.5\times10^{19}$ cm$^{-3}$, and $4\times10^{19}$ cm$^{-3}$. Other unintentional dopants, such as nitrogen and oxygen, are also measured and found to have low concentrations, which are not expected to significantly affect the thermal conductivity. The depth profiles of the respective dopants for all four sample sets are presented in Figure S1 of the Supplementary Materials (SM). The sample thicknesses are measured using the picosecond acoustic technique in the TDTR system and the result is shown in Fig. 1A, with the inset illustrating the principle. The detailed principle of thin film thickness measurement by picosecond acoustic technique can be found in Materials and Methods section. The measured thickness (~3.5 µm) is much larger than the thermal penetration depth in the TDTR measurements (further explanation see SM text), which supports that the samples are thermally thick and can be treated as bulk crystals. The large-area EBSD maps presented in Figs. S2 and S3 demonstrate the single (111) orientation of both the unintentionally doped and boron-doped ($6\times10^{18}$ cm$^{-3}$) 3C-SiC samples. The absence of grain boundaries in these maps further confirms the high crystalline quality and monocrystalline nature of the samples.

Figure 1B shows the atomic-scale high-angle annular dark field (HAADF)-STEM image of the high-quality 3C-SiC crystal, with the interplanar spacings of the (111) and ($11\bar{1}$) lattice planes agree well with references(*22, 23*). The inset image shows the fast Fourier transform (FFT) of the HAADF-STEM image, indicating the single-crystal nature of the selected area. Large-area TEM and STEM images of the unintentionally and heavily boron-doped ($1.5\times10^{19}$ cm$^{-3}$) samples are provided in Figure S4 for more comprehensive structural analysis. In Fig. S4A, the folds observed in both the 3C-SiC layer and the Si substrate are artifacts caused by the FIB sample, rather than intrinsic material defects. Figure S4B clearly reveals the 3C-SiC/Si heterointerface, exhibiting a high density of threading dislocations and structural defects resulting from the substantial lattice mismatch between 3C-SiC and Si(*24*). However, due to the limited thermal penetration depth, our thermal measurements are not sensitive to the low-quality interfacial regions of the sample. Figures S4C and S4D show the HAADF-STEM images of the atomic crystal structure of the two samples in a larger area and their corresponding FFT images.

To further verify the sample's crystalline quality, we performed Raman spectroscopy analysis, which provides sensitive detection of structural defects and strain in the material. The Raman spectra of both the 3C-SiC samples in this work (upper curve) and a free-standing 3C-SiC bulk sample (lower curve)(*17*) are shown in Fig. 1C. The shifts and full width at half maximum (FWHM) values of the Raman peaks of both samples are consistent, indicating the high quality of the films. The broadening at the base of the second Raman peak is attributed to the Si substrate(*25*). The longitudinal optical (LO) and transverse optical (TO) phonon frequencies at $\Gamma$ point which correspond to the Raman peak positions are illustrated in Fig. 1D. Good agreement is observed with the calculated 3C-SiC phonon dispersion curve(*26*).

The spatial distribution of B dopants was analyzed using APT, which is capable of mapping the three-dimensional distribution of specific elements at atomic resolution(*27, 28*). The sample for APT measurements is fabricated into a needle-shaped tip using a focused ion beam (FIB) system to enhance the localized electric field near the tip, thereby facilitating the evaporation of surface atoms (Fig. 2A). More details about APT measurements can be found in the Materials and Methods section. The atomic maps of silicon and carbon are shown in Figs. 2B and 2C, respectively. These two elements, which form the host material

SiC, are uniformly distributed. The atomic map of B (Fig. 2D) reveals a wave-like distribution at the atomic scale, different from the uniform distributions of silicon and carbon. However, no B clustering is observed, and SIMS measurements show uniform distributions at a larger scale as shown in Fig. S1. Therefore, we expect that the non-uniform B distribution has minimal impact on the thermal conductivity of 3C-SiC. The mass spectrum acquired after adjacent averaging smoothing via APT is shown in Fig. 2E, with the three elements highlighted in different colors. Variations in the mass-to-charge peaks of the same element arise due to the presence of isotopes and doubly charged ions. The original data without smoothing is provided in Fig. S5. In Fig. S6 the comprehensive distribution of all three elements is shown. It is notable that this should be the first atomic scale characterization data about B doping in 3C-SiC.

**Thermal characterization**

The thermal conductivity of the 3C-SiC single crystals was measured by the TDTR technique. More description and details about TDTR measurement can be found in Materials and Methods section, SM text, Figs. S7-S8, and Ref(*29*). In Fig. S8, the sensitivities of the Al/SiC thermal boundary conductance (TBC) and the SiC thermal conductivity are high, with distinct shapes over the delay times, allowing them to be fitted with high accuracy. Figure 3A presents the TDTR fitting data of the four samples with different doping levels. As the B doping concentration increases, the TDTR ratio decreases, indicating a reduction in the thermal conductivity of 3C-SiC single crystals. The mechanism behind this strong reduction in thermal conductivity is illustrated in Fig. 3B. When B atoms substitute for C atoms, the bond lengths between the defect and surrounding Si atoms undergo slight modifications, leading to a transition from tetrahedral to threefold symmetry(*14, 15*). This transition introduces a strong perturbation in the interatomic force constants (IFC), as exemplified by the one-dimensional potential energy curve shown in Fig. 3B. The geometric distortion of the central atom causes the overlap of two potential minima, resulting in a significant change in the second derivative of the potential energy, which directly influences the IFC(*20*). This strong perturbation induces an anomalous peak in the phonon-defect scattering rate at low frequency. Thus, the thermal conductivity was dramatically suppressed because low frequency acoustic phonons carry most of the heat. As a result, phonons near this resonant frequency experience significantly stronger scattering with B defects, leading to a much larger thermal conductivity reduction than that predicted by mass differences alone.

Figure 3C compares our experimental results with previous experimental and theoretical studies. The excellent agreement between the measured and calculated thermal conductivity of the undoped single crystal shows that our samples have high-quality and are thermally thick in the TDTR measurements. Moreover, the measured thermal conductivity of the doped crystals also aligns well with theoretical predictions, providing strong evidence for resonant phonon scattering induced by intentional B doping. This scattering phenomenon significantly affects the thermal conductivity of 3C-SiC, leading to an approximately 50% reduction at a doping concentration of $\sim 4\times 10^{19}$ cm$^{-3}$. Additionally, the theoretical thermal conductivities for varying N concentrations, as predicted by Katre *et al*(*15*). and Pang *et al*(*20*)., are also shown in Fig. 3C. At a doping concentration of $1\times 10^{19}$ cm$^{-3}$, N impurity has a negligible effect on thermal conductivity, whereas B doping results in a thermal conductivity reduction of approximately 25%. The experimentally measured thermal conductivity of nitrogen-doped 3C-SiC single crystals is also included in Fig. 3C(*30*) for comparison, which agrees well with theoretical predictions. To achieve a similar reduction in thermal conductivity of 3C-SiC, the nitrogen doping concentrations need to be tens of times in magnitude higher than the B doping. This occurs because the scattering mechanisms in the N-doped 3C-SiC are mostly mass difference-induced phonon-defect scattering and phonon-electron scattering (in Pang *et al.*(*20*) and Wang *et al.*(*26*)), without the contribution of significant resonant phonon scattering. If the symmetry transition effect were ignored for B-doped 3C-SiC, the impact on thermal conductivity would be close to that of Al atoms substituting for Si atoms, corresponding to the effect of largely mass difference-induced scattering. The orange solid line in Fig. 3C represents this scenario, showing only a ~7% reduction in thermal conductivity at a doping concentration of $1\times10^{19}$ cm$^{-3}$, which is significantly weaker than the B-doped case.

This study reports the experimental observation of the strongest thermal conductivity reduction caused by B doping in common semiconductor materials. Figure 3D shows the normalized thermal conductivity reduction in various semiconductors doped with different impurities, including commonly used semiconductors across a broad range of bandgaps, such as silicon(*31*), germanium(*32*), gallium arsenide(*33*, *34*), silicon carbide(*30*, *35*), gallium nitride(*36*, *37*), aluminum nitride(*37–40*), and diamond(*41*). The doped elements for each material are also shown. The solid lines represent theoretical predictions for silicon carbide(*20*) doped with nitrogen, which is in good consistence with experiment(*30*). It is clearly evident that B-doped 3C-SiC exhibits the largest percentage reduction in thermal conductivity across the entire studied concentration range.

We also notice that there are large variations and controversy for the experimental data in the literature. Most of the samples were not comprehensively characterized to obtain the detailed information such as crystal quality, dopant distributions, and other structural imperfections. Potential contributions from grain boundaries, dislocations, and unintentional defects—along with their spatial distributions—could also contribute to the thermal conductivity reduction, but comprehensive studies characterizing and analyzing these factors are currently unavailable in the literature. Moreover, systematic and accurate thermal conductivity measurements under controlled conditions are still lacking. These make it difficult to fully capture the impact of impurity doping alone on thermal conductivity independently, underscoring the importance of a thorough examination of doping concentration and dopant atom distribution in this study.

Figure 4A compares the measured thermal conductivity of B-doped 3C-SiC from this study with literature data for N and B co-doped 3C-SiC reported by Ivanova *et al.*(*19*), heavily N-doped 3*C*-SiC from Huang *et al.*(*30*), and DFT calculations from Katre *et al.*(*15*). The unintentionally doped 3C-SiC single crystal has a thermal conductivity that excellently matches the DFT-calculated thermal conductivity value of perfect 3C-SiC single crystal, highlighting again the high quality of the samples in this work. Also, it shows that the effects of other unintentionally doped impurities on the thermal conductivity of the 3C-SiC single crystals in this study are negligible. In comparison, the co-doped polycrystalline 3C-SiC ceramic samples from Ivanova *et al.*(*19*) exhibit much lower thermal conductivity, attributed to intense phonon scattering at grain boundaries and by other structural imperfections.

To systematically investigate thermal transport in B-doped 3C-SiC, we also measured the temperature dependent thermal conductivity of the four samples from 300 to 700 K, as shown in Figs. 4B and 4C. Figure 4B presents the temperature-dependent thermal conductivities for samples with the lowest and highest doping concentrations. The results of unintentional doped 3C-SiC shows excellent agreement with the DFT prediction of Katre *et al.*(*15*) over the entire temperature range. Figure 4C compares our experimental results of the other two doping levels with the theoretical predictions of Pang *et al.*(*20*). At similar doping concentrations ($1.5\times10^{19}$ cm$^{-3}$ and $0.6\times10^{19}$ cm$^{-3}$ versus $1.0\times10^{19}$ cm$^{-3}$), our measurements align well with the theoretical predictions up to 500K. All measured data follow logarithmic linearity, converging at elevated temperatures as Umklapp phonon-phonon scatterings dominate over phonon-defect scatterings(*42*). Additionally, as the doping concentration increases, the temperature dependence of thermal conductivity weakens, as reflected in a reduced absolute slope in logarithmic scale, the $T^{-1}$ line is also shown as reference in Figs. 4B and 4C as a baseline. This trend indicates that phonon-defect scattering progressively dominates over phonon-phonon scattering at higher doping levels at low to medium temperature.

**Discussion**

In summary, this study presents the achievement and experimental observation of extremely strong defect-phonon scatterings in B-doped single crystal 3C-SiC, showcasing the strongest thermal conductivity reduction—nearly 50% at a doping concentration of $4\times10^{19}$ cm$^{-3}$—observed among common semiconductors. Using the advanced TDTR method, we systematically measured the thermal conductivities across different doping levels, with results in excellent agreement with resonant phonon scattering theory. Moreover, temperature-dependent measurements further confirm room temperature observations. Advanced structural characterizations, including HAADF-STEM, Raman spectroscopy, and EBSD, confirmed the high quality of the 3C-SiC single crystal and enabled the exclusion of other structural imperfections from B doping. SIMS and APT provided detailed analyses of the nanoscale scale dopant concentration and the atomic scale spatial distribution—collectively reinforcing the critical role of B impurities. This work advances the fundamental understanding of phonon scattering mechanisms in semiconductors and provides critical guidance for thermal management of next-generation semiconductor devices.

**Materials and Methods**

**Sample growth and preparation**

The single-crystal 3C-SiC samples studied in this work are grown by Air Water Inc. The growth process is performed on a (111) Si substrate at 1300 K via LT-CVD in a custom-designed CVD reactor. Both the growth temperature and Si substrate orientation are carefully optimized to enable high-quality single-crystalline 3C-SiC epitaxial growth. These 3C-SiC single crystals are expected to have low crystal defect densities including stacking faults and double positioning boundaries at relatively low growth temperature due to the shared rotational symmetry (120°) between Si and 3C-SiC around the [111] axis. The 3C-SiC films are then polished to a final thickness of ~3.5 μm (thermally thick regime) with smooth surface for TDTR measurements.

**Thermal characterizations**

The thermal conductivity of the 3C-SiC crystals was measured using TDTR, a non-contact optical technique based on a femtosecond laser source and pump-probe interaction. In TDTR, the pump beam is modulated at a specific frequency and focused onto the sample through an objective lens, inducing periodic temperature fluctuations. Meanwhile, a mechanical delay stage adjusts the optical path of the pump beam to control the delay time between the arrivals of the pump and probe beams. The probe beam, focused by the same objective lens, detects surface temperature variations by thermoreflectance. To enhance the thermoreflectance signal, an ~85 nm aluminum (Al) transducer layer was deposited on the sample surface. The reflected probe beam is detected by a silicon photodiode, and a lock-in amplifier extracts the signal component at the modulation frequency. By fitting the measured signal to an analytical solution of the sample structure, the unknown thermal properties—specifically, the thermal conductivity of 3C-SiC and the TBC of the Al/3C-SiC interface—can be determined. In this study, the pump and probe beam radii were both 12 μm, with a modulation frequency of 10.1 MHz. Further details are provided in the SM text, Figs S7 and S8.

**Picosecond acoustic technique**

An ultrafast laser pulse train generates coherent strain waves that propagate through the thin-film layers at the longitudinal phonon group velocity, reflecting independently at each material interface. The reflected strain waves modulate the optical reflectivity of the Al transducer surface, which is subsequently detected by the probe beam at progressive time delays. These modulations appear as peaks or valleys in the in-phase signal, determined by the acoustic impedance $Z$ contrast between adjacent materials at each interface. Here $Z = \rho v$, where $\rho$ is the density of the layer and $v$ is the sound speed. The round-trip propagation time of strain waves through individual layers is obtained from the temporal separation between corresponding echo features, more details about can be found in SM text. Using this approach, the thickness of each film can be precisely determined at the local measurement spot.

**Raman spectroscopy**

Raman spectroscopy was performed using a Horiba LabRAM confocal Raman imaging system. The acquisition time is 60 s over 5 cycles. The laser wavelength is 532 nm. The objective is 50x, and the total energy setting is 5%.

**SIMS characterizations**

The precise depth profiles of doping elements, including B, N, and O, within the crystals were determined using SIMS. A device from Cameca Ltd. accelerated $C_s^+$ to 10keV to detect N and O, and a beam of $O_2^+$ accelerated at 12.5keV was used to measure the concentration of B. The SIMS characterizations was conducted at Evans Material Technology (Shanghai) Co., Ltd.

**TEM and STEM measurements**

TEM and STEM samples were prepared by a Thermo-Fisher Helios G4 UX system with a milling accelerating voltage of 30 kV. An accelerating voltage of 2kV is applied to remove the amorphous damage layer. Pt protective layer is deposited onto the milling zone in advance. The TEM and HAADF-STEM images of the 3C-SiC crystal structure and 3C-SiC/Si interface are acquired by a Thermo-Fisher FEI Titan Cubed Themis G2 300 at an accelerating voltage of 300 kV.

**EBSD measurement**

The crystal orientation maps of the sample surfaces are obtained using EBSD measurement. A high-resolution Hitachi SU5000 scanning electron microscope (SEM) generates backscattered electrons containing crystallographic information, which are detected by an EDAX Hikari CCD camera. The scanning process is performed over a total area of 500 μm×500 μm with a step size of 3 μm. The EBSD measurements are conducted at WinTech Nano-Technology Services Pte Ltd.

**APT characterization**

The atom probe tomography (APT) specimens were prepared by Ga FIB milling. The APT measurements were performed with a local electrode atom probe (LEAP 5000XR) under UV laser (355 nm) pulsing at pulse energy of 80 pJ, a pulse repetition rate of 200 kHz and a target evaporation rate of 0.5% per pulse at 40 K. The reconstruction and quantitative analysis of APT data were performed using a CAMECA visualization and analysis software (AP Suite 6.3). The APT measurement is conducted at Shanghai Fuxin Technology Consulting Co., Ltd.

**Acknowledgments**

Z.H., Y.W, and Z.C. thank Prof. Bo Sun for allowing them to use his TDTR system. The authors acknowledge Electron Microscopy Laboratory of Peking University, China for the use of Cs corrected Titan Cubed Themis G2 300 transmission electron microscopy. **Funding:** Z. H., Y.W, and Z.C. acknowledge the funding support from "National Key Research and Development Program of China" under the Grant No. 2024YFA1207900, and "The Fundamental Research Funds for the Central Universities, Peking University". **Author contributions:** Conceptualization: Z. H., J. L. and Z.C. Data curation: Z. H., Y.W, and Z.C. Funding acquisition: J. L., M. L., R. W., and Z. C. Investigation: Z. H., Y.W, and Z.C. Methodology: Z. H., Y.W, and Z.C. Project administration: Z. C. Resources: L. J. and Z. C. Visualization: Z. H. and Z.C. Supervision: Z. S., N. S., M. L., R. W., and Z. C.

Writing—original draft: Z. H. and Z. C. Writing – review & editing: Z. H., J. L., Y. W., and Z. C. **Competing interests:** Authors declare that they have no competing interests.

**Data and materials availability:** All data are available in the main text or the supplementary materials.

**Figures and Tables**

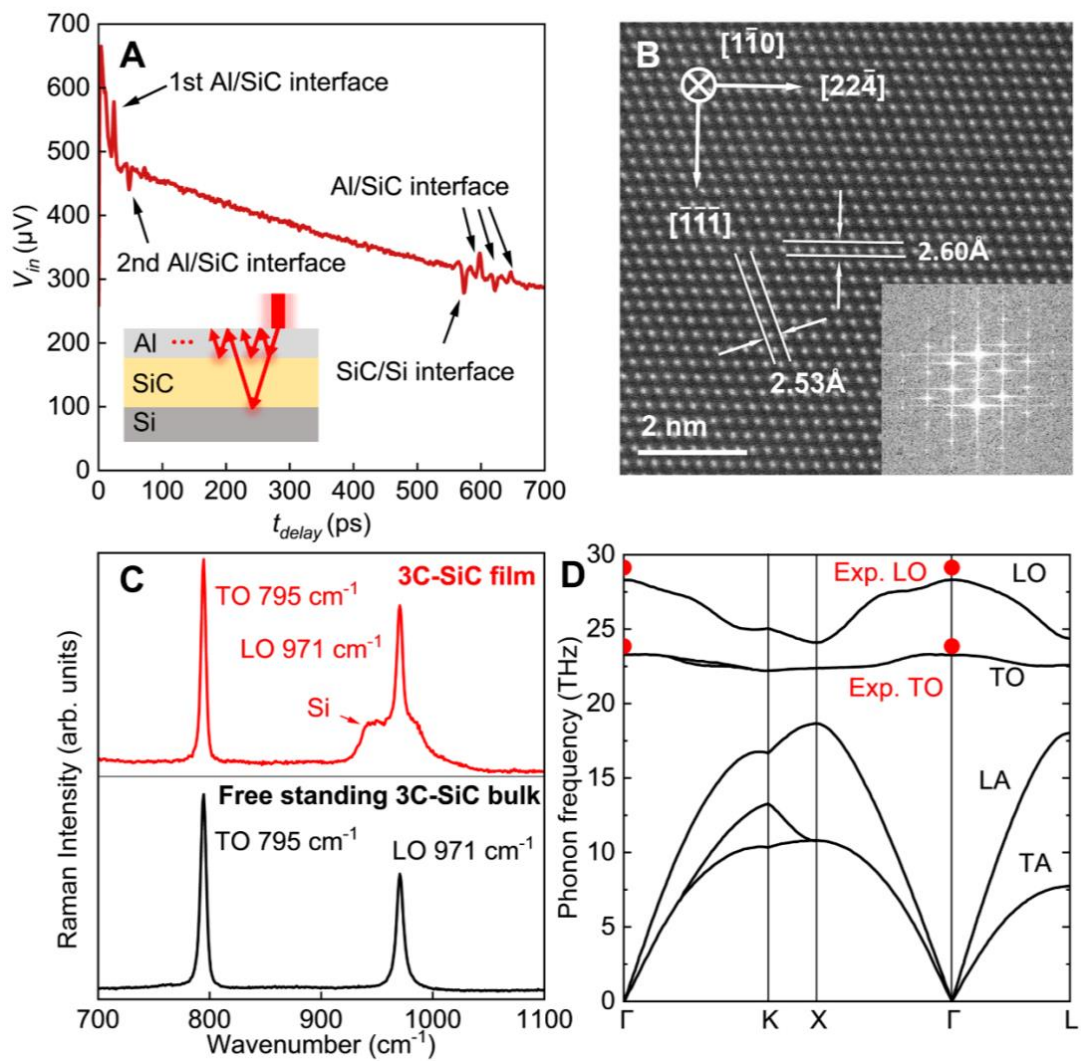

**Fig. 1. Structural characterizations of the 3C-SiC sample.** (**A**) Picosecond acoustic signal of the sample, with the inset illustrating the measurement principle. (**B**) HAADF-STEM image of the 3C-SiC crystal layer along [1$\bar{1}$0] axis, with the inset showing the FFT of the STEM image. (**C**) Raman spectra of the 3C-SiC film in this study (upper curve) and a free-standing 3C-SiC bulk (lower curve). The broadening at the base of the peak is attributed to the Si substrate(*25*). (**D**) The TO and LO phonon frequencies at Γ point correspond to the Raman results, comparing with the calculated phonon dispersion curve(*26*).

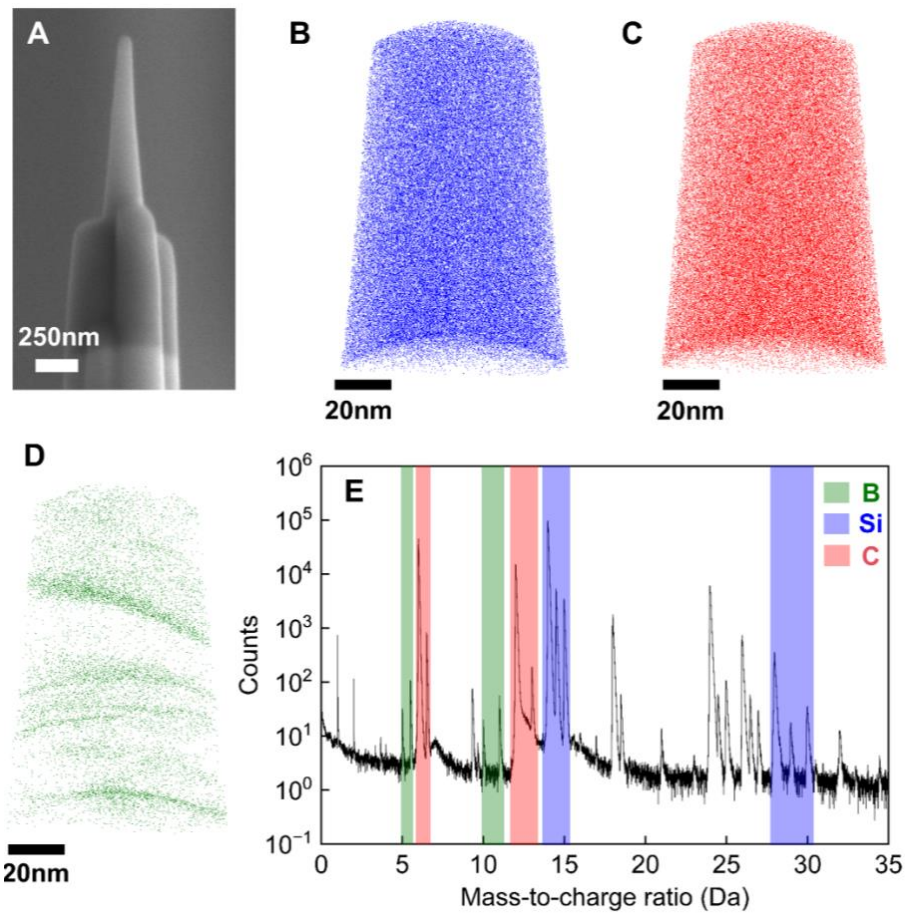

**Fig. 2. Atomic-scale element distributions in 3C-SiC revealed by APT.** (**A**) Needle-shaped sample prepared for APT characterization. (**B-D**) Atomic distribution maps of silicon (**B**), carbon (**C**), and boron (**D**). (**E**) Mass spectrum acquired via APT, with peak positions corresponding to boron, carbon, and boron.

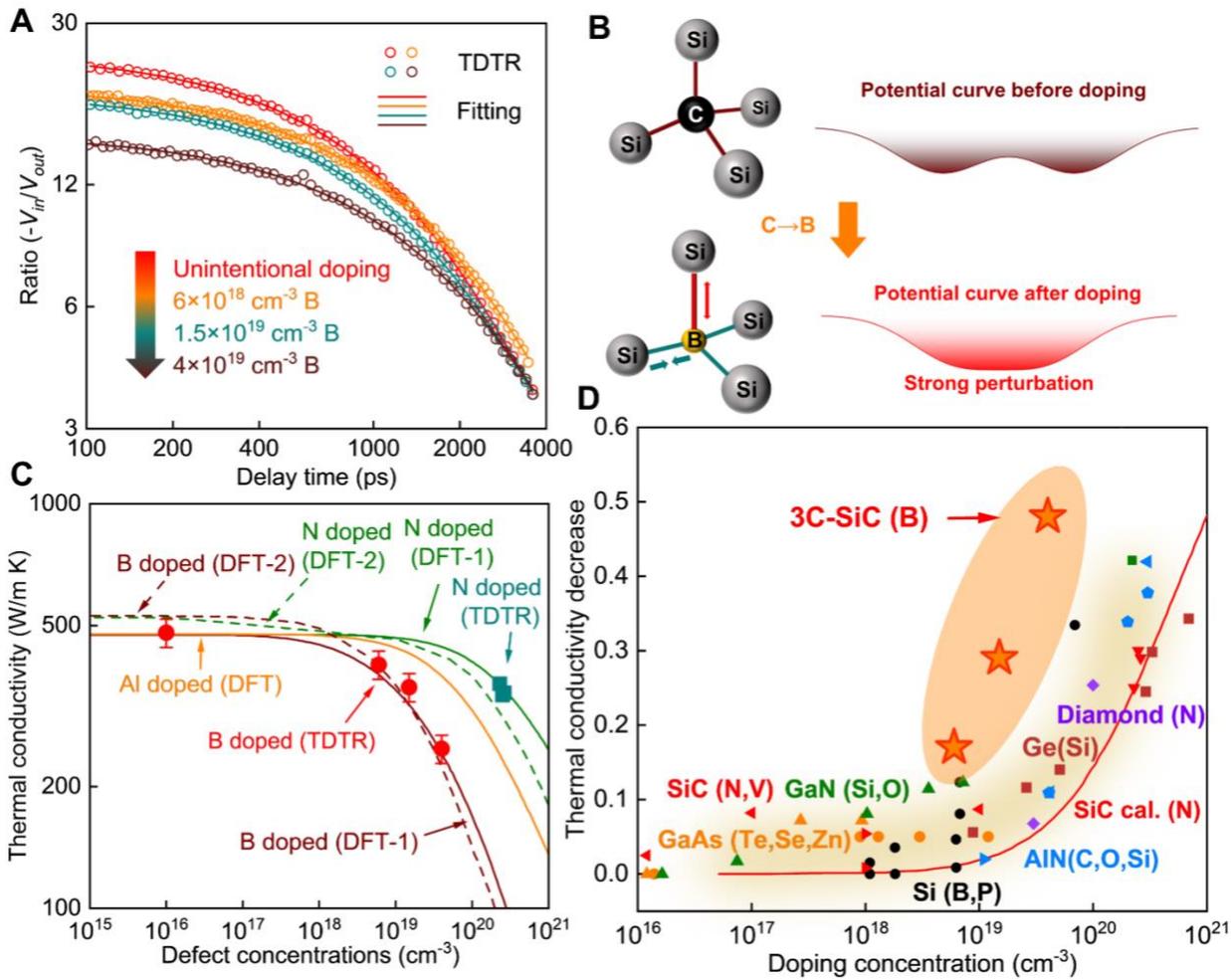

**Fig. 3. Thermal transport in 3C-SiC single crystals at different boron doping concentrations.** (**A**) TDTR experimental data and corresponding best-fit results for different doping concentrations. (**B**) Schematic illustration of resonant phonon scattering around a B defect in 3C-SiC: the substitution of a B atom for a C atom induces structural asymmetry, significantly perturbing the local potential field. (**C**) Comparison of experimental and theoretical studies on the relationship between thermal conductivity and doping concentrations in 3C-SiC. (**D**) Summary of normalized thermal conductivity reductions in different common semiconductors under various doping conditions(*20,30-41*).

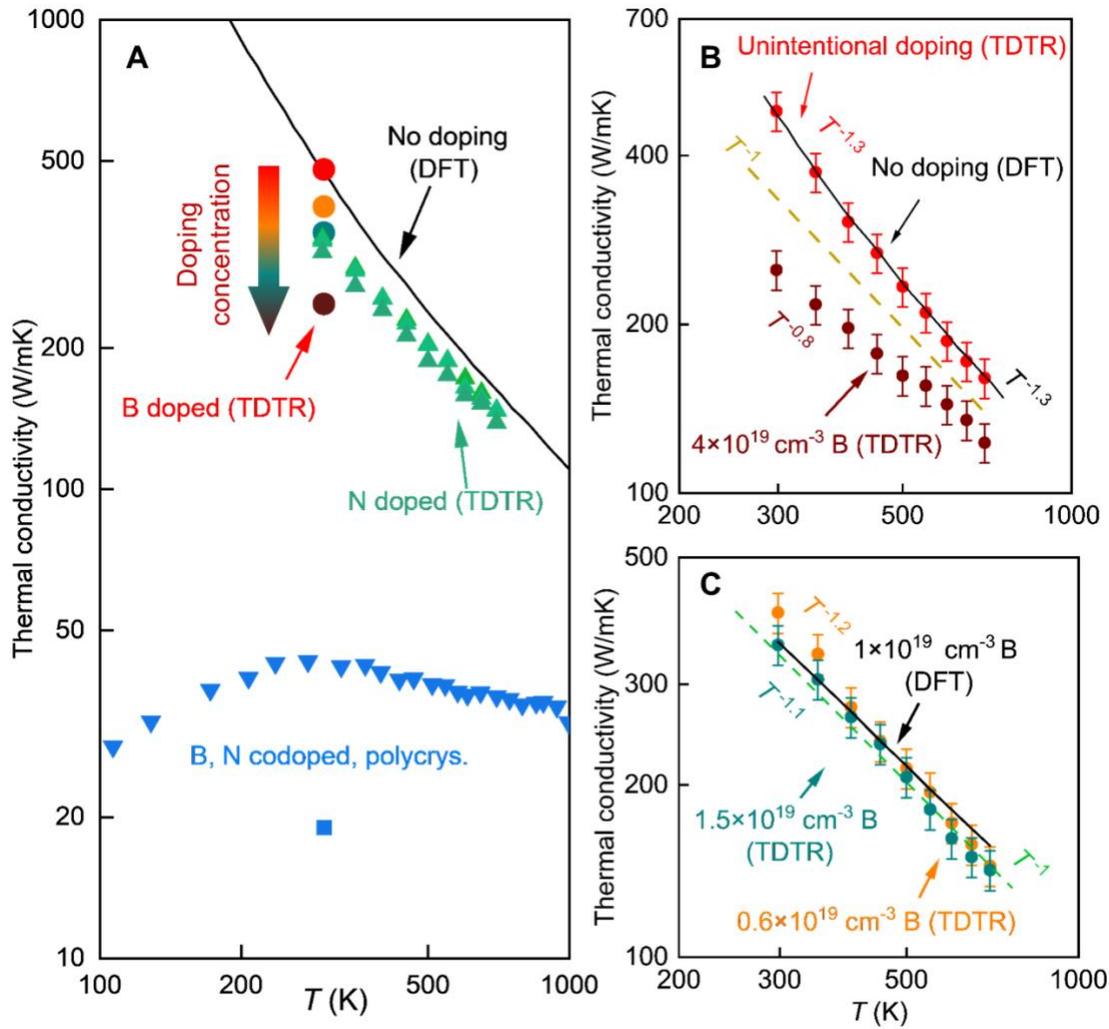

**Fig. 4. Temperature dependence of thermal conduction in 3C-SiC single crystals.** (**A**) Comparison of the measured thermal conductivity in this study with literature data from Ivanova et al.(*19*) and Huang et al.(*30*) alongside DFT calculations from Katre et al.(*15*). (**B**) Temperature-dependent thermal conductivity of unintentionally doped and $4\times10^{19}$ cm$^{-3}$ B-doped 3C-SiC, compared with the calculation results from Katre et al.(*15*). (**C**) Temperature-dependent thermal conductivity of 3C-SiC doped with $6\times10^{18}$ cm$^{-3}$ and $1.5\times10^{19}$ cm$^{-3}$ B, alongside the theoretical prediction of $1\times10^{19}$ cm$^{-3}$ B-doped from Pang et al.(*20*).

**Supplementary Materials**

**This PDF file includes:**

    Supplementary Text

    Figs. S1 to S8

    References (*43-44*)